\documentclass[journal,twoside,print]{ieeecolor}

\usepackage{tmi_arxive}
\usepackage{cite}
\usepackage{amsmath,amssymb,amsfonts}
\usepackage{algorithmic}
\usepackage{graphicx}
\usepackage{textcomp}
\usepackage{placeins}
\usepackage{caption}
\usepackage{cuted}

\def\BibTeX{{\rm B\kern-.05em{\sc i\kern-.025em b}\kern-.08em
    T\kern-.1667em\lower.7ex\hbox{E}\kern-.125emX}}

\begin{document}

\title{The First Nozzle-Mounted Compton Camera Prompt Gamma Imaging System for In Vivo Proton Therapy Dose Verification}

\author{
Farshad Safavi, \IEEEmembership{Member, IEEE},
Stephen W. Peterson,
Sina Mossahebi,
Ananta Chalise,
Vijay R. Sharma,
Matthias K. Gobbert,
Jerimy C. Polf,
Lei Ren%
\thanks{Farshad Safavi, Sina Mossahebi, Ananta Chalise, and Lei Ren are with the Department of Radiation Oncology, University of Maryland School of Medicine, Baltimore, MD, USA.}
\thanks{Stephen W. Peterson is with the Department of Physics, University of Cape Town, Cape Town, South Africa.}
\thanks{Vijay R. Sharma is with the Department of Medical Physics, University of Wisconsin School of Medicine and Public Health, Madison, WI, USA.}
\thanks{Matthias K. Gobbert is with the Department of Mathematics and Statistics, University of Maryland, Baltimore County, Baltimore, MD, USA.}
\thanks{Jerimy C. Polf is with H3D, Inc., Ann Arbor, MI, USA.}
}

\maketitle

\begin{abstract}
This study presents the first clinical integration and experimental demonstration of a nozzle-mounted Compton camera prompt gamma imaging (PGI) system for \textit{in vivo} proton
range verification. Four position-sensitive solid-state Compton
camera modules, each containing four cadmium zinc telluride
(CdZnTe) detector crystals, were integrated into a modified
range shifter mounted directly on the treatment nozzle of a
clinical proton therapy gantry. This compact fixed-geometry
configuration maintained alignment with the proton beam axis
throughout irradiation and enabled stable synchronized data
acquisition during pencil-beam scanning delivery. The system was evaluated under realistic clinical proton beam
delivery conditions using single-energy and spread-out Bragg
peak (SOBP) irradiations at gantry angles of 90° and 270°, delivered doses of 2~Gy and 7.5~Gy, and controlled distal range shifts of up to 10~mm. Prompt gamma events were reconstructed into three-dimensional emission distributions using a physics-based Compton scatter reconstruction framework. The system
operated reliably during all irradiations and produced reproducible
prompt-gamma localization across repeated measurements.
Reconstructed emission distributions remained geometrically
consistent across gantry angles and demonstrated sensitivity to
controlled distal range perturbations, with measurable upstream
shifts of the emission hotspot corresponding to reduced proton
penetration depth. These results demonstrate the feasibility of a clinically integrated
nozzle-mounted quad-camera Compton PGI system for detecting
millimeter-scale proton range variations during beam delivery and
represent an important step toward clinically deployable prompt
gamma--based \textit{in vivo} treatment verification in proton therapy.
\end{abstract}

\maketitle

\begin{figure}[!t]
    \centering
    \includegraphics[width=\linewidth]{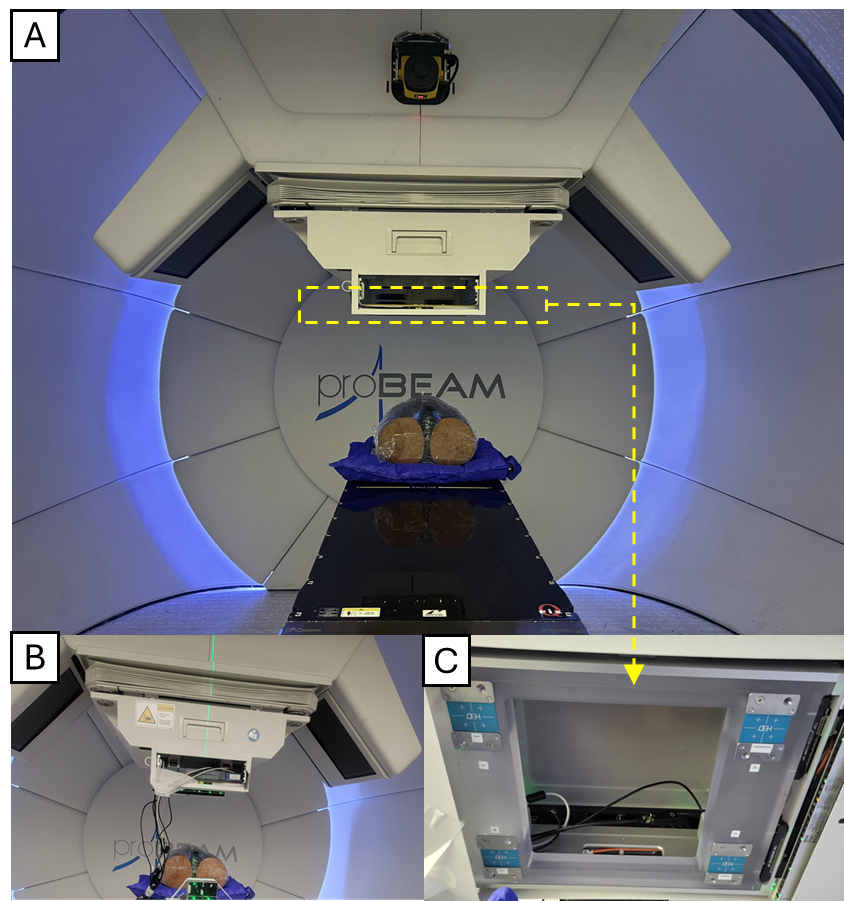}
    \caption{ (A) Nozzle-mounted Prompt Gamma Imaging (PGI) system has been integrated on a clinical proton therapy gantry for in-vivo dose verification. The PGI assembly is mounted beneath the treatment nozzle, aligned with the proton beam axis and positioned above the anthropomorphic phantom during irradiation. (B) Experimental configuration of the nozzle-mounted PGI system aligned with the proton beam axis during irradiation of an anthropomorphic phantom. (C) Close-up view of the detector assembly showing the solid-state Compton camera modules integrated at the corners of the modified range shifter for prompt-gamma detection.}
    \label{fig:system}
\end{figure}


\section{Introduction}
\label{sec:introduction}

\IEEEPARstart{P}{roton} therapy enables highly conformal dose delivery due to the finite penetration range of charged particles and the sharp distal falloff at the Bragg peak, which allows precise dose delivery while sparing surrounding healthy tissues beyond the target depth \cite{ref1,ref2}. However, the precision of proton therapy treatments has been limited by the proton beam range uncertainties arising from anatomical variations, patient setup errors, and inaccuracies in CT-based stopping-power ratio (SPR) estimation. To mitigate range uncertainties, safety treatment margins of approximately ±3.5\% are commonly incorporated into treatment planning \cite{ref4}. Although these margins ensure adequate tumor coverage, the resulting larger treatment volume leads to increased dose and toxicities in healthy tissues and limits our ability for dose escalation to enhance the tumor control \cite{ref3,3_1}. Reliable in-vivo verification of proton beam range during treatment delivery therefore represents an important unmet clinical need. 

Several approaches for treatment verification have been investigated. Positron emission tomography (PET) imaging has been investigated for treatment verification immediately after the proton therapy delivery. This approach was limited by the time-sensitivity of the signal and not providing verification during treatment, which are both critical to guide treatment intervention to minimize the impact of range uncertainties. To address this limitation, substantial research efforts have focused on real-time proton range verification during beam delivery \cite{ref5,ref6,ref7}. Among these, PGI has emerged as a promising method since prompt gamma rays are emitted nearly instantaneously via proton–nuclear interactions and provide a direct surrogate for the proton beam range \cite{polf2009prompt}. 

Early clinical implementations of prompt gamma in-vivo range verification demonstrated successful integration of PG monitoring into proton therapy workflows and confirmed its capability for treatment verification during patient irradiation \cite{ref11, ref12}. Initial PG imaging systems using slit collimation enabled real-time monitoring but suffered from limited field of view, low detection efficiency, and one-dimensional range information \cite{vijay}. To overcome these limitations, Compton camera–based PGI has attracted considerable attention due to its capability to reconstruct three-dimensional (3D) gamma emission distributions without mechanical collimation. Early feasibility studies demonstrated successful imaging of prompt gamma emissions during the delivery of clinical proton beams using Compton cameras, establishing their potential for proton range verification \cite{ref_anchor2}. Subsequent investigations demonstrated reconstruction of 3D prompt gamma emission distributions under clinically realistic irradiation conditions and detection of proton range shifts on the order of 2–3 mm \cite{ref_anchor3}.

In parallel, significant effort has focused on improving image reconstruction algorithms for Compton camera–based PG imaging. Comparative studies evaluating Maximum Likelihood Expectation Maximization (MLEM) and Stochastic Origin Ensemble (SOE) reconstruction methods demonstrated millimeter-level agreement between reconstructed prompt-gamma profiles and proton dose falloff \cite{ref8}. Kernel-based reconstruction approaches such as the Kernel Weighted Back Projection (KWBP) algorithm further improved reconstruction robustness by reducing noise and enhancing spatial localization accuracy in phantom experiments \cite{ref13}. 

Despite these advances, most previously reported PGI systems rely on single-camera configurations, limiting spatial coverage and reconstruction capability \cite{XIE2017210}. In contrast, the proposed nozzle-mounted quad-camera configuration increases the effective solid-angle coverage around the proton beam axis, improving prompt-gamma detection efficiency and enabling more robust three-dimensional localization of emission distributions during clinical beam delivery.

This work presents a novel clinical integration of a nozzle-mounted quad-camera Compton PGI system installed directly on a clinical proton therapy gantry, as illustrated in Fig.~\ref{fig:system}. The imaging assembly is mounted on the treatment nozzle and aligned with the proton beam axis, enabling prompt gamma acquisition during irradiation under realistic clinical delivery conditions.  An overview of the proposed quad-camera detector configuration is shown in Fig.~\ref{fig:system}. The main contributions of this work are summarized as follows:


\begin{itemize}

\item Development and clinical deployment of, to the best of our knowledge, the first nozzle-mounted quad-Compton camera PGI system for in-vivo proton therapy dose verification. Unlike previous Compton camera PGI systems positioned in the treatment room outside the beam nozzle, which remain fixed in the room coordinate system, the proposed nozzle-mounted system rotates with the proton beam as the gantry rotates.

\item Design and implementation of a compact fixed-geometry range shifter with integrating Compton camera detector modules as shown in Fig.~\ref{fig:system}(C), allowing minimizing disruption to the routine clinical workflow.

\item Development of a real-time data acquisition and processing pipeline enabling prompt-gamma reconstruction using kernel-weighted back projection.

\item Experimental validation of a prompt-gamma–based proton range verification system under clinically relevant beam energies, gantry angles, and controlled range-shift conditions.

\item Quantitative evaluation of system feasibility and reconstruction reproducibility for in-vivo proton beam range verification.

\end{itemize}

The remainder of this paper is organized as follows. 
Section~\ref{sec:Methods} describes the proposed prompt-gamma imaging system, including the clinical overview of the PGI system, the architecture of the Compton quad camera detector and the integration of the system with the proton therapy gantry. 
Section~\ref{sec:Results} presents experimental validation results that include prompt-gamma detection and three-dimensional (3D) reconstruction, Comparison of SOBP beam delivery at gantry angles $90^\circ$ and $270^\circ$, and sensitivity to controlled distal range changes. 
Finally, Section~\ref{sec:Conclusion} summarizes the findings and discusses future directions for clinical implementation of prompt-gamma–based proton range verification.

\FloatBarrier
\section{Methods}
\label{sec:Methods}

\begin{figure*}[t]
    \centering
    \includegraphics[width=\textwidth]{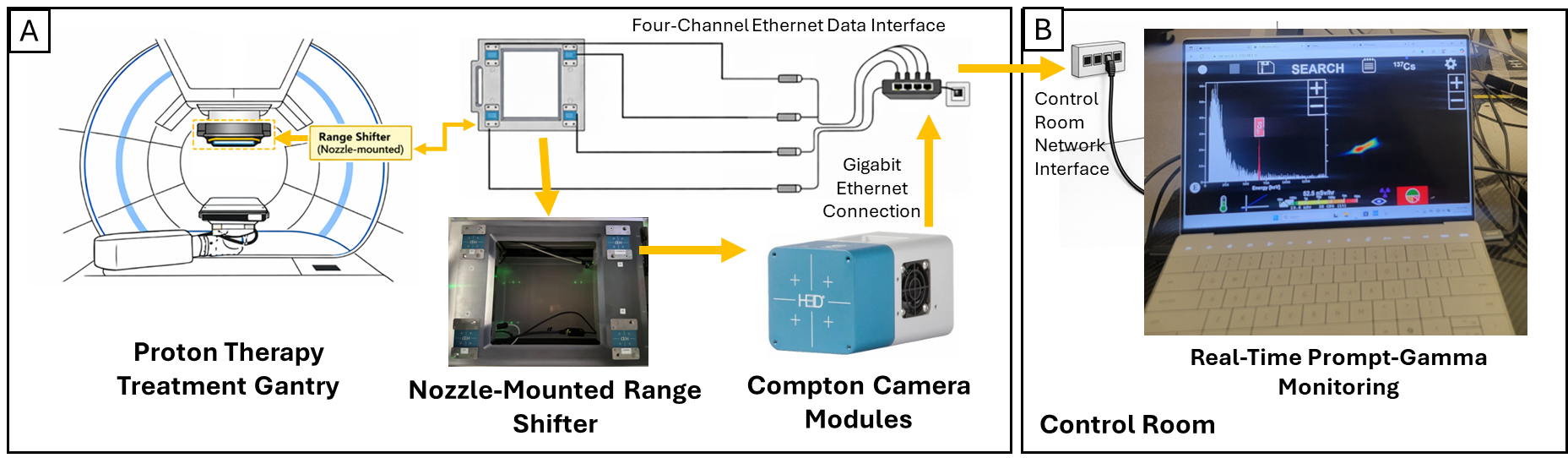}
    \caption{
    (A) Schematic overview of the quad-camera Compton imaging system integrated with the nozzle-mounted range shifter and connected via a four-channel Gigabit Ethernet interface for data acquisition.
    (B) Real-time prompt-gamma monitoring interface displayed in the control room during proton beam delivery.
    }
    \label{fig:figure3}
\end{figure*}

\subsection{Clinical PGI Overview}
\begin{figure*}[t]
    \centering
    \includegraphics[width=\textwidth]{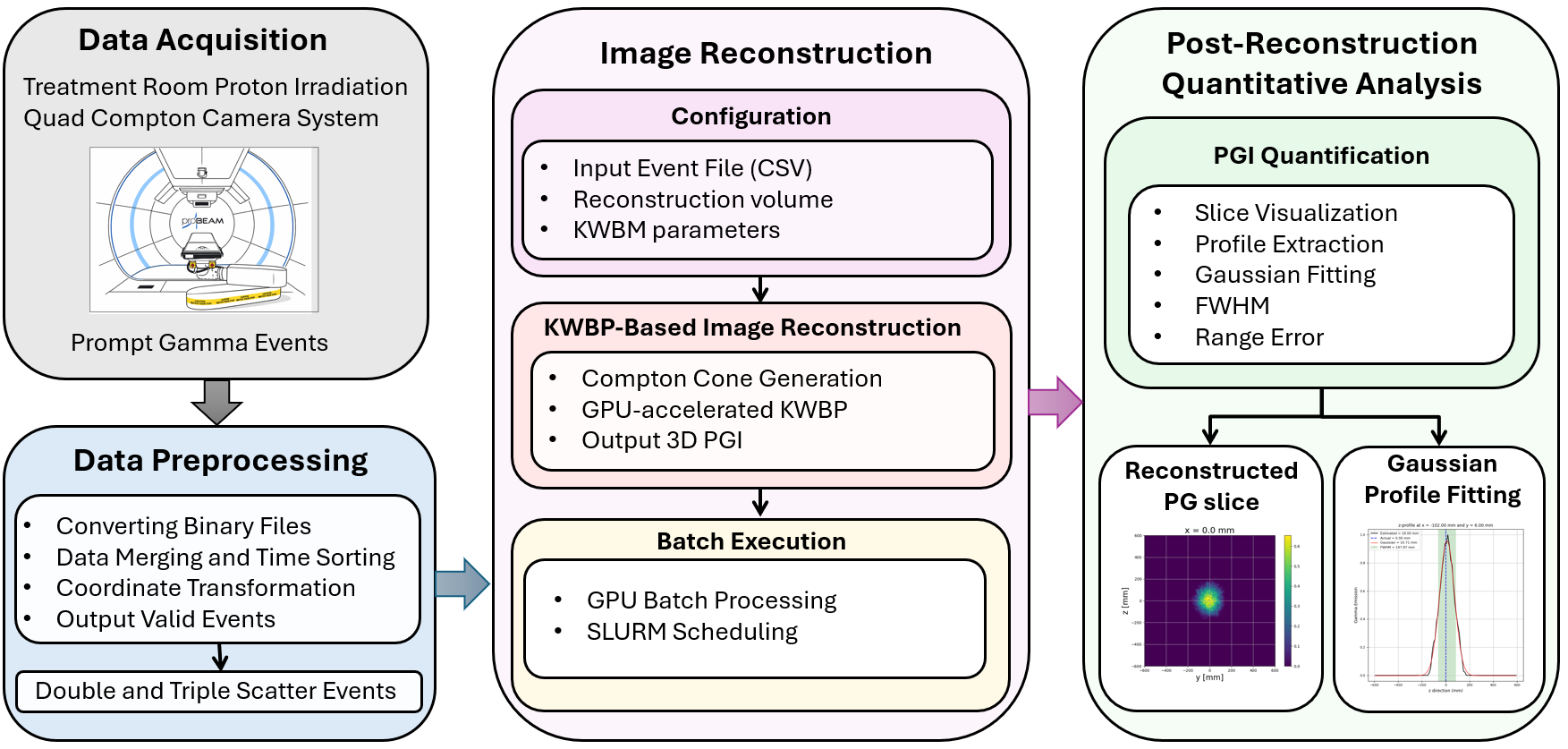}
    \caption{ PGI reconstruction framework. Prompt-gamma events acquired during proton irradiation are preprocessed through data merging, time sorting, coordinate transformation, and event filtering. The filtered events are reconstructed into a three-dimensional prompt-gamma image using a GPU-accelerated Kernel Weighted Backprojection (KWBP) algorithm. Post-reconstruction analysis includes slice visualization, profile extraction, Gaussian fitting, full width at half maximum (FWHM) estimation, and localization error evaluation.}
    \label{fig:pgi_framework}
\end{figure*}

A PGI system was developed for in-vivo proton range verification through clinical integration of a nozzle-mounted quad-camera Compton imaging configuration, as shown in Fig.~\ref{fig:figure3}. The system consists of four position-sensitive CdZnTe Compton camera modules integrated within a modified range shifter mounted on the proton therapy nozzle, enabling fixed alignment with the proton beam axis during gantry rotation. Prompt-gamma interactions generated during irradiation are acquired in list-mode and transmitted to the control room via a multi-channel Gigabit Ethernet interface for real-time capture, as depicted in Fig.~\ref{fig:figure3}. The acquired data are subsequently reconstructed into 3D prompt-gamma emission distributions, illustrated in Fig.~\ref{fig:pgi_framework}, enabling visualization of beam-related emission profiles and facilitating proton range verification in the patient coordinate system.

\subsection{Quad Compton Camera PGI System}

The PGI system employs four position-sensitive CdZnTe Compton camera modules arranged in a quad-detector configuration surrounding the proton beam axis, as shown in  Fig.~\ref{fig:system}. Each compton camera module has compact dimensions of approximately $10.2 \times 5.7 \times 5.7$~cm$^3$ contains pixelated CdZnTe crystals providing a total sensitive volume exceeding 19~cm$^3$. The detectors operate at room temperature and provide high-resolution gamma spectroscopy suitable for prompt-gamma detection in a clinical proton therapy environment.

The Compton cameras provide an energy resolution of 0.8\% FWHM at 662~keV \cite{H3D_M400_specs} and record details of each interaction, including the deposited energy, the position of the interaction, and the timestamp information. These capabilities enable accurate identification of the multiple gamma interactions required for Compton imaging reconstruction. Each module supports real-time event readout by streaming the binary interaction data via Ethernet connection, allowing synchronized acquisition across multiple detector units, as shown in Fig.~\ref{fig:figure3}(A).

To integrate the detectors within the clinical proton therapy environment, a custom mechanical assembly was developed by modifying a Lucite range shifter mounted on the proton therapy gantry snout, as shown in Fig.~\ref{fig:system}(A). A central opening of $18.6 \times 18.6$~cm$^2$ was machined into a 4~cm-thick Lucite slab to allow unobstructed transmission of the proton beam. Four additional $5.7 \times 5.7$~cm$^2$ cavities were created at the corners of the range shifter to house the Compton camera modules. The center of each detector was positioned approximately 5~cm from the edge of the central aperture, resulting in a symmetric quad-camera arrangement surrounding the beam axis.

The quad-camera detector configuration is integrated into the range
shifter, enabling direct mounting on the treatment nozzle to ensure compatibility with clinical proton beam delivery. For treatment fields smaller than the $18.6 \times 18.6$~cm$^2$ aperture at the level of the range shifter, the presence of the detector holder does not interfere with proton spot delivery. The nozzle-mounted design ensures fixed detector alignment relative to the proton beam axis during gantry rotation, providing stable system geometry for PGI under clinical treatment conditions. 

The proposed configuration introduces a novel approach for clinical PGI. Unlike previously reported PGI systems, the proposed system embeds the Compton camera modules directly within the range-shifter structure mounted on the treatment nozzle. This design provides a compact and mechanically stable geometry that preserves standard clinical beam delivery while enabling prompt-gamma acquisition during irradiation without introducing additional structures into the treatment room. By integrating the detectors into an existing beamline component, the system supports practical deployment of PGI in a clinical proton therapy environment.

To enable remote operation in the clinical environment, the detector system was connected from the treatment room to the control room through a Gigabit Ethernet interface. Prompt-gamma event data acquired by the nozzle-mounted quad-camera system were transmitted over Ethernet to a laptop in the control room, enabling monitoring during proton beam delivery. The real-time prompt-gamma monitoring interface displayed in the control room is shown in Fig.~\ref{fig:figure3}(B). This configuration supported practical data acquisition in the treatment setting while keeping the monitoring workstation outside the treatment room.

\subsection{Prompt Gamma Image Reconstruction}

Prompt gamma data acquired during proton irradiation were processed using a modular reconstruction framework designed for clinical integration and reproducible evaluation. The general processing pipeline, including data acquisition, reconstruction and post-reconstruction quantitative analysis, is illustrated in Fig.~\ref{fig:pgi_framework}. The workflow converts detector interaction data into 3D prompt-gamma emission distributions through Compton imaging reconstruction and quantitative post-processing.

\subsubsection{Data Acquisition and Preprocessing}

Raw binary data from the four Compton camera modules were first converted to text format, then merged into a single time-sorted dataset, with unit conversion and basic interaction filtering applied to generate the combined event file used for reconstruction. Interaction positions recorded in detector local coordinates were then transformed into the global patient coordinate system using gantry dependent rigid body transformations defined by module-specific rotation matrices and translation vectors derived from the quad-camera geometry. Finally, physics-based event filtering was applied to identify valid double and triple scatter interactions prior to image reconstruction. The filtering criteria included energy windowing, Compton scatter consistency checks, scattering-angle limits, interaction separation constraints, and distance of closest approach (DCA) requirements to ensure physically consistent Compton events.

\subsubsection{Prompt Gamma Image Reconstruction}

The PGI reconstruction used in this study is illustrated in Fig.~\ref{fig:pgi_framework} (Image Reconstruction). The reconstruction framework consists of three main components: configuration,  Kernel Weighted Backprojection (KWBP), and batch execution.

\textbf{Configuration:}
In the configuration stage, reconstruction parameters and event-selection criteria were specified. These parameters include the selection of valid double and triple scatter events, the dimensions and resolution of the reconstruction grid, and algorithm-specific settings such as smoothing and noise sampling kernel parameters used by the KWBP reconstruction algorithm.

\textbf{Kernel Weighted Backprojection Image Reconstruction:}
Prompt-gamma image reconstruction was performed using a kernel weighted backprojection approach. For each detected prompt-gamma event, the measured energy deposits and interaction positions define a Compton cone that constrains the possible origin of the gamma ray. The Compton scattering angle $\theta$ is determined from the measured energies according to the Compton scattering equation

\begin{equation}
\cos\theta =
1 - m_e c^2
\left(
\frac{1}{E_2} - \frac{1}{E_1 + E_2}
\right),
\end{equation}

where $E_1$ and $E_2$ denote the energy deposits at the first and second interaction sites and $m_e c^2 = 511$ keV is the electron rest energy.

Each Compton cone contributes to the reconstruction volume through a kernel-weighted backprojection process. The contributions from all detected events are accumulated over the voxelized reconstruction grid to generate a three-dimensional prompt-gamma emission distribution. GPU-based parallel computation was used to accelerate the backprojection and enable efficient processing of large event datasets.

\textbf{Batch Execution:} To support large experimental datasets and repeated reconstruction runs, the reconstruction pipeline was executed using an automated batch-processing framework within a high-performance GPU-based computing environment, enabling efficient processing across multiple datasets and reconstruction configurations.

\FloatBarrier
\section{Results}
\label{sec:Results}
\begin{figure*}[h]
    \centering
    \includegraphics[width=\textwidth]{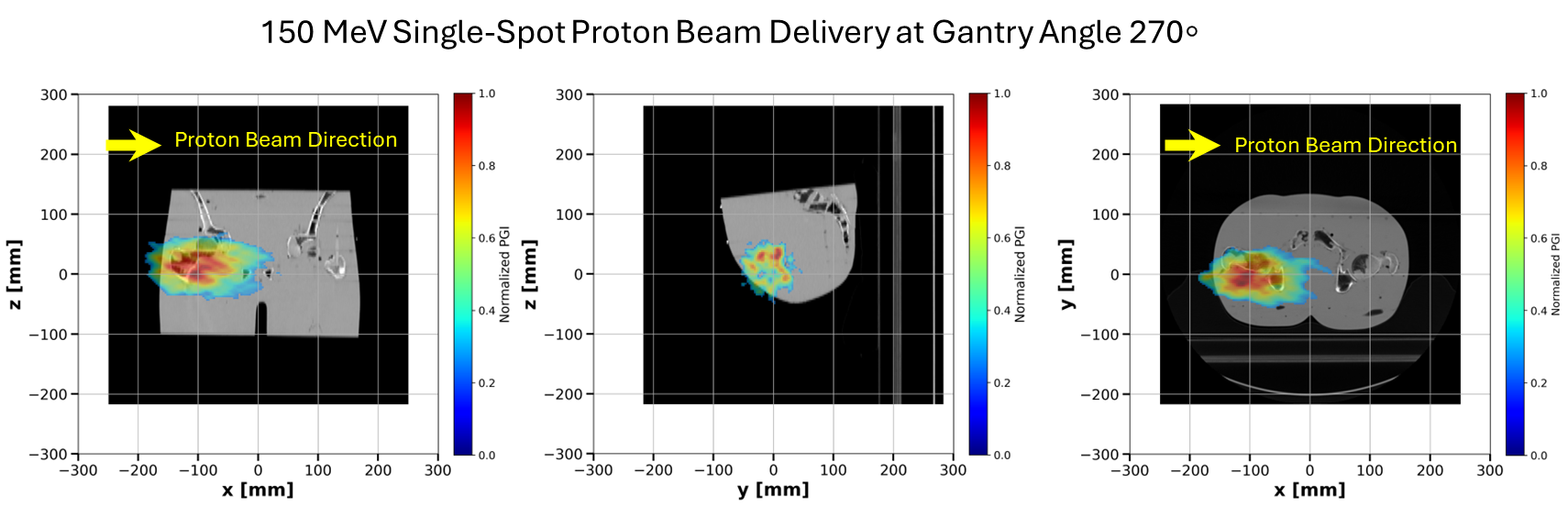}
    \caption{Validation of prompt-gamma detection during clinical proton beam delivery. Reconstructed 3D prompt-gamma emission distributions for a 150 MeV single-spot proton beam delivered at a gantry angle of 270°. Orthogonal slices (x–z, y–z, and x–y) of the reconstructed volume show localization of the emission region along the beam direction, demonstrating successful detection and reconstruction of prompt-gamma events using the nozzle-mounted quad-camera PGI system.}
    \label{fig:detection}
\end{figure*}

\begin{figure}[t]
    \centering
    \includegraphics[width=\linewidth]{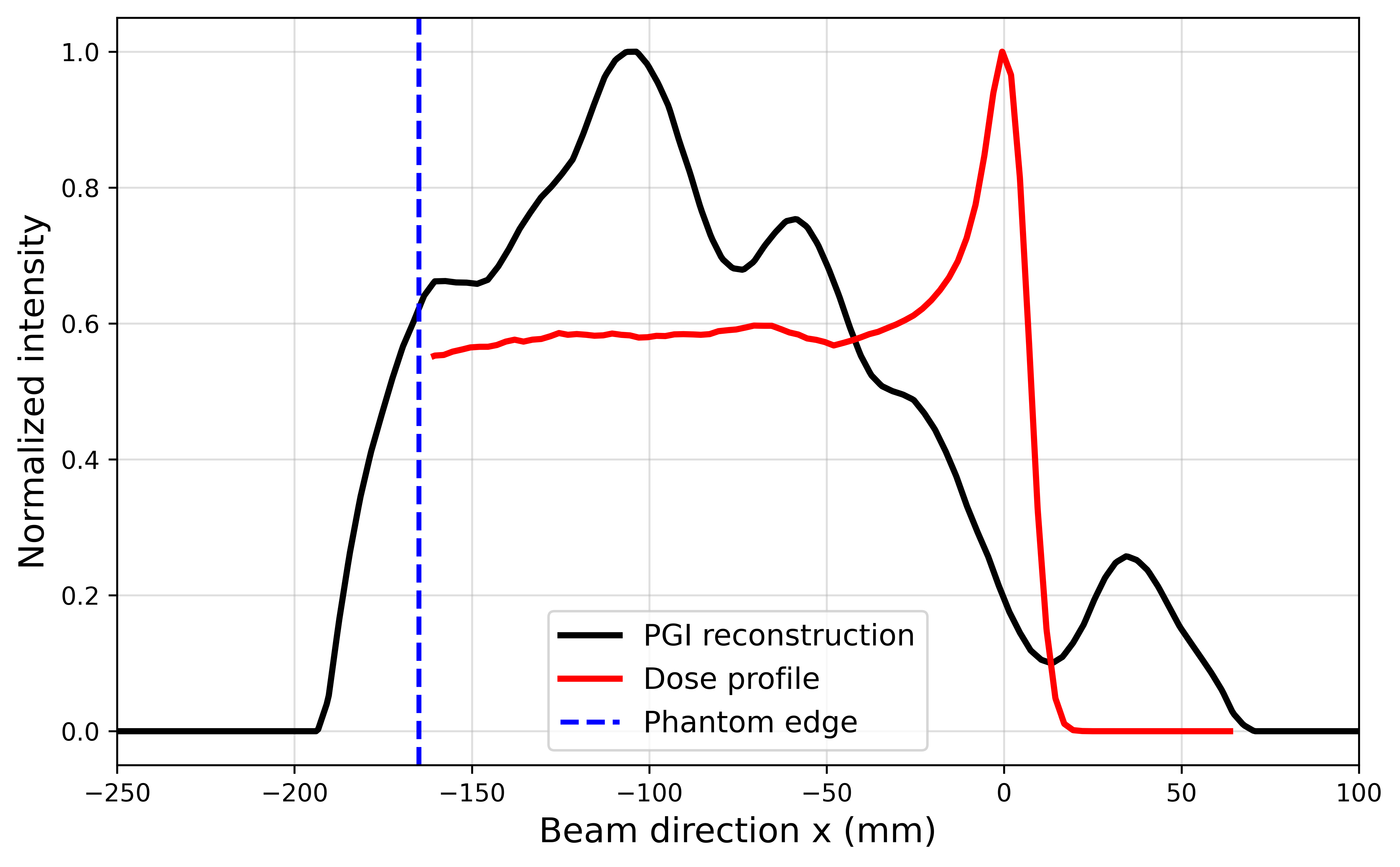}
    \caption{
    Normalized prompt-gamma and proton dose profiles along the beam direction extracted at y = 0 mm and z = 0 mm. The black curve represents the reconstructed prompt-gamma emission profile obtained from the quad-camera Compton imaging system, while the red curve shows the proton dose profile from the treatment planning system (Bragg peak). The dashed blue line indicates the distal edge of the phantom.
    }
    \label{fig:profile}
\end{figure}

\begin{figure}[t]
    \centering
    \includegraphics[width=\linewidth]{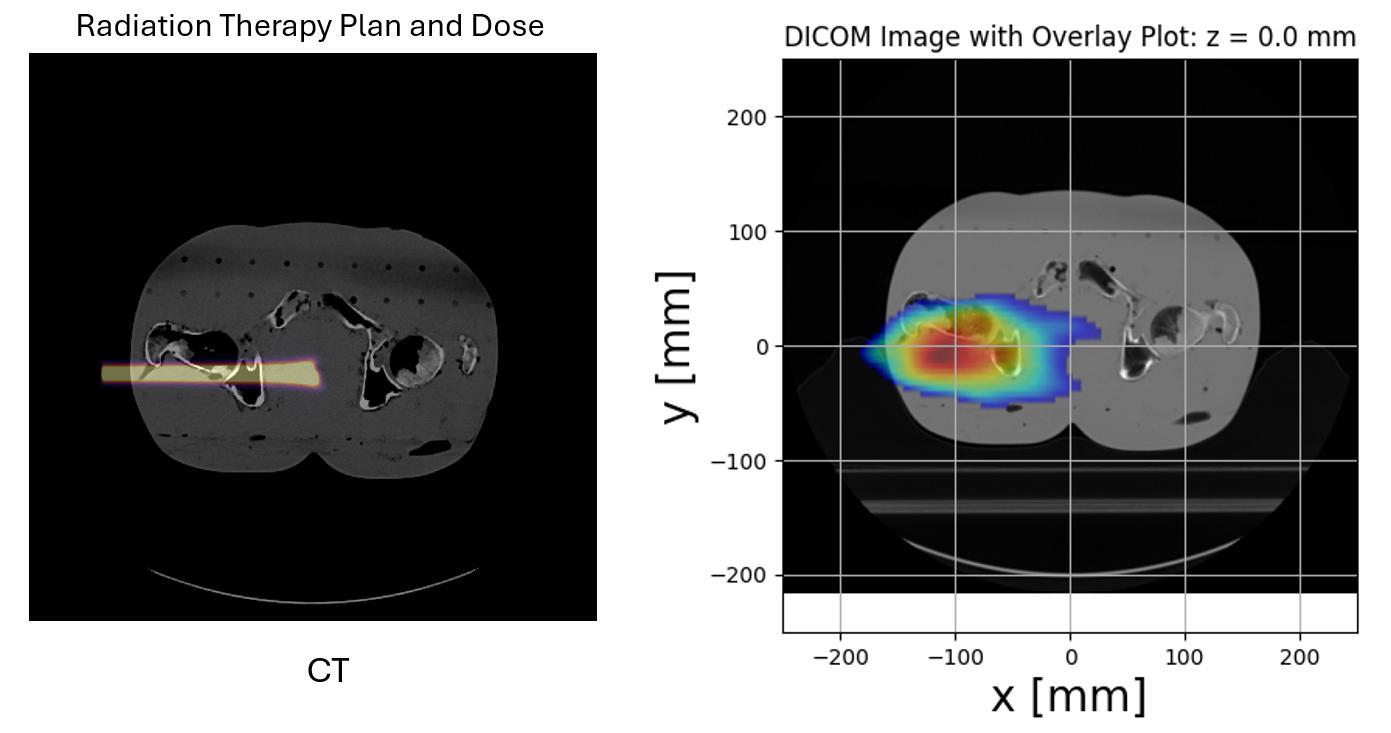}
    \caption{
     CT-registered prompt-gamma emission distributions shown in axial view. (Left) Planning CT with proton beam path and dose distribution. (Right) Reconstructed prompt-gamma emission overlaid on the planning CT, demonstrating localization of the emission region along the proton beam path.}
    \label{fig:overlay_profile}
\end{figure}

\begin{figure*}[t]
\centering
\includegraphics[width=0.85\textwidth]{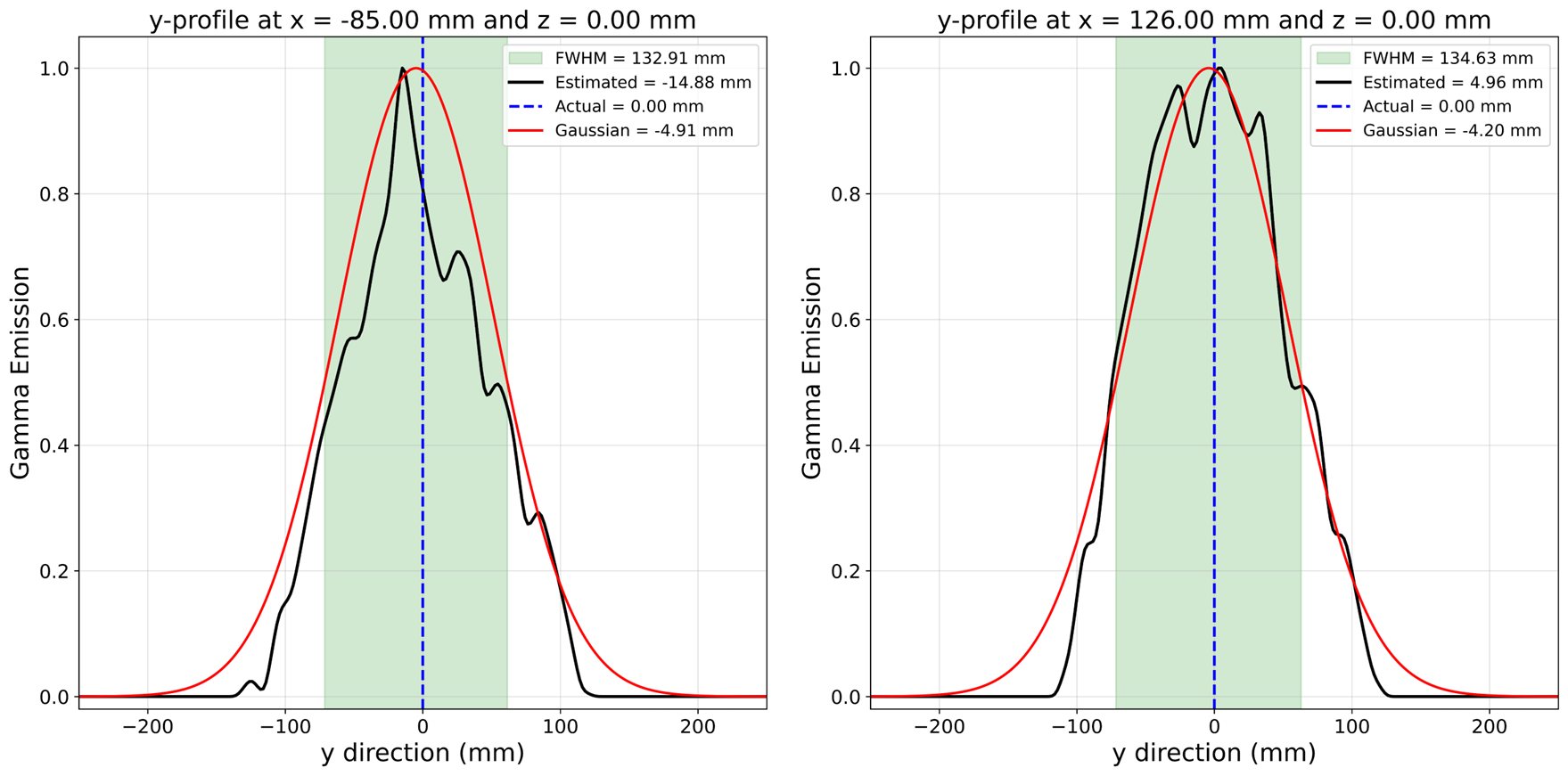}
\caption{
Lateral prompt-gamma emission profiles reconstructed at gantry angles $90^\circ$ and $270^\circ$ during delivery of a single-spot proton beam. 
\textit{Left:} Profile extracted for gantry angle $90^\circ$ at $x=-85.0$ mm and $z=0.0$ mm. 
\textit{Right:} Profile extracted for gantry angle $270^\circ$ at $x=126.0$ mm and $z=0.0$ mm. 
Measured prompt-gamma emission distributions (black curves) are shown together with Gaussian fits (red curves), expected beam positions (blue dashed lines), and the corresponding FWHM regions (green shaded areas). 
Despite detector rotation with gantry angle, consistent peak localization is observed, demonstrating stable prompt-gamma detection and reliable geometric reconstruction across different treatment orientations.
}
\label{fig:gantry_profile_comparison}
\end{figure*}

This section presents the experimental validation of the proposed
nozzle-mounted quad-camera PGI system using an anthropomorphic
phantom under realistic clinical proton beam delivery conditions.
All measurements were performed during phantom irradiation in the
treatment room to evaluate the operational stability and imaging
performance of the system in a clinical treatment environment.

System performance was evaluated through three experimental studies.
First, prompt-gamma detection and 3D emission
reconstruction were demonstrated during single-spot proton beam
delivery, verifying the ability of the system to localize the emission
region along the beam path. Second, measurements were acquired at gantry angles of $90^\circ$ and $270^\circ$ to evaluate the effect of gantry rotation on the reconstructed prompt-gamma emission profiles. Third,
controlled distal range perturbations of up to 10~mm were introduced to assess the sensitivity of the reconstructed prompt-gamma distributions to beam range changes.
Across all experiments, the system demonstrated stable operation during irradiation without interfering with the treatment beam delivery. The
reconstructed prompt-gamma distributions showed reproducible 3D localization of the emission region, while
quantitative profile analysis confirmed sensitivity to millimeter-scale
distal range shifts. These results support the feasibility of the
proposed configuration for in-vivo proton range verification.

\subsection{Prompt Gamma Detection and 3D Reconstruction}

Figures~\ref{fig:detection} 
and~\ref{fig:profile} demonstrate successful detection and reconstruction of prompt-gamma emission during clinical proton beam delivery using the nozzle-mounted quad-camera Compton imaging system. Figure~\ref{fig:detection} shows the reconstructed 3D prompt-gamma emission distribution obtained during delivery of a 150~MeV single-spot proton beam at a gantry angle of $270^\circ$, with a beam current of $80$~nA and a total delivery of $60$~kMU. Orthogonal slices of the reconstructed volume are shown in the coronal ($x$--$z$), sagittal ($y$--$z$), and axial ($x$--$y$) planes, corresponding to standard anatomical imaging views used in clinical computed tomography (CT). The reconstructed emission region is clearly localized along the proton beam trajectory, with the prompt-gamma intensity concentrated near the expected beam path. The consistency of the reconstructed hotspot across all three orthogonal views confirms accurate three-dimensional localization of prompt-gamma events and demonstrates reliable detection and reconstruction within the imaging field of view under clinical beam delivery conditions. This multi-planar visualization enables direct comparison with clinical CT images and supports the feasibility of using the proposed nozzle-mounted prompt-gamma imaging system for in-room verification of proton beam range during patient treatment.

To further quantify the reconstructed emission distribution, a lateral intensity profile was extracted from the reconstructed volume along the $x$-direction at $y = 0$~mm and $z = 0$~mm (Fig.~\ref{fig:profile}). The measured prompt-gamma emission profile was fitted with a Gaussian model to estimate the centroid and spatial spread of the reconstructed emission region. The Gaussian centroid provides an estimate of the reconstructed beam location, while the FWHM characterizes the lateral spread of the emission distribution. The peak of the depth emission profile does not line up directly with the Bragg Peak location due to the detector configuration. The nozzle view of the detectors magnifies the signal from prompts gammas produced at the entrance portion of the beam thus producing the emission peak shown in Fig.~\ref{fig:profile} near the surface of the phantom.  This effect has been studied and can be corrected by modeling the gamma attenuation as discussed in \cite{vijay}.

\begin{figure*}[t]
\centering
\includegraphics[width=\textwidth]{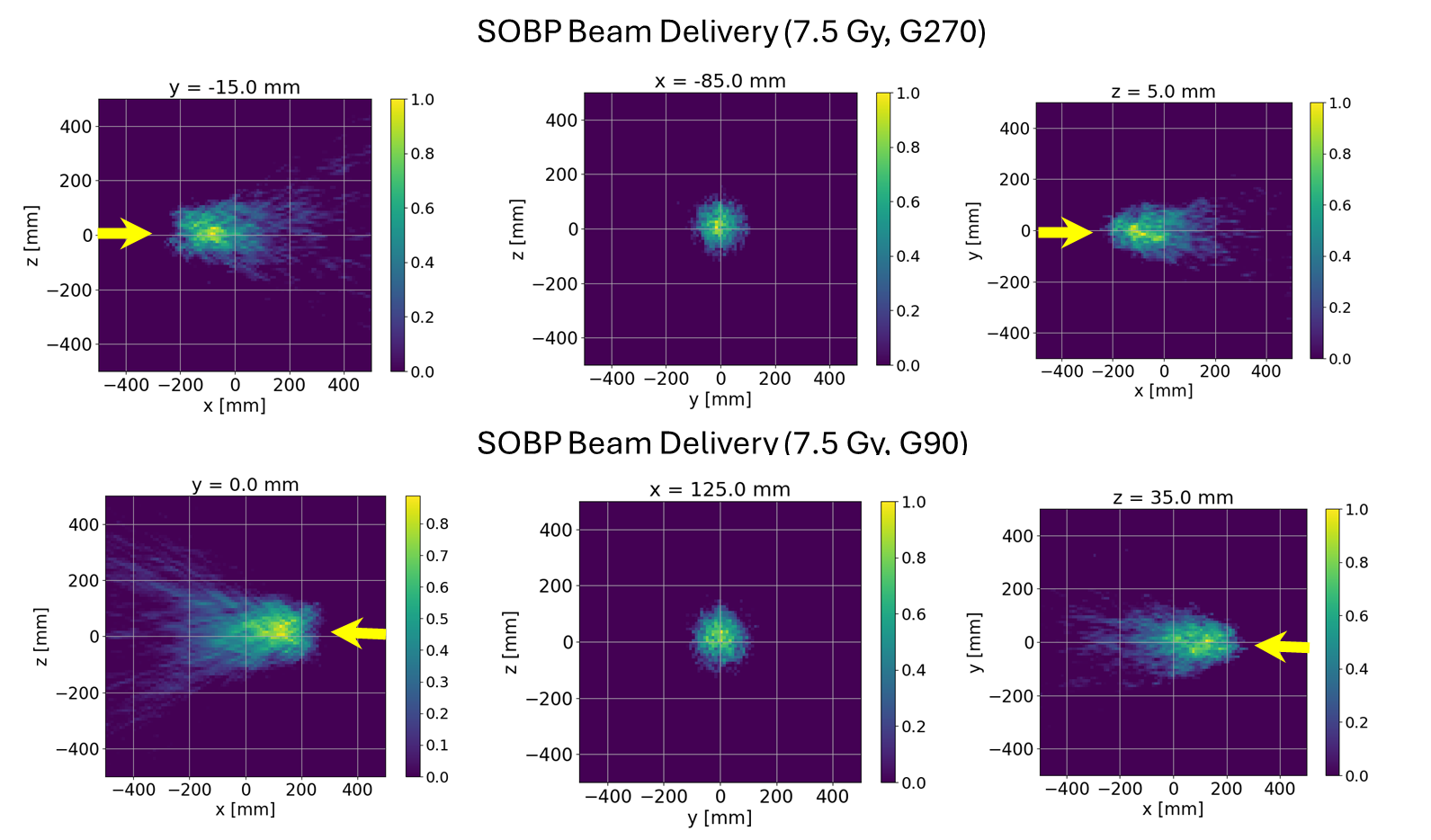}
\caption{ Comparison of reconstructed prompt-gamma emission for a 7.5~Gy SOBP beam at G270 (top) and G90 (bottom). Emission elongation aligns with the beam
direction while lateral localization remains stable across gantry rotation.
}
\label{fig:AngleEffect}
\end{figure*}

\begin{figure}[t]
\centering
\includegraphics[width=\columnwidth]{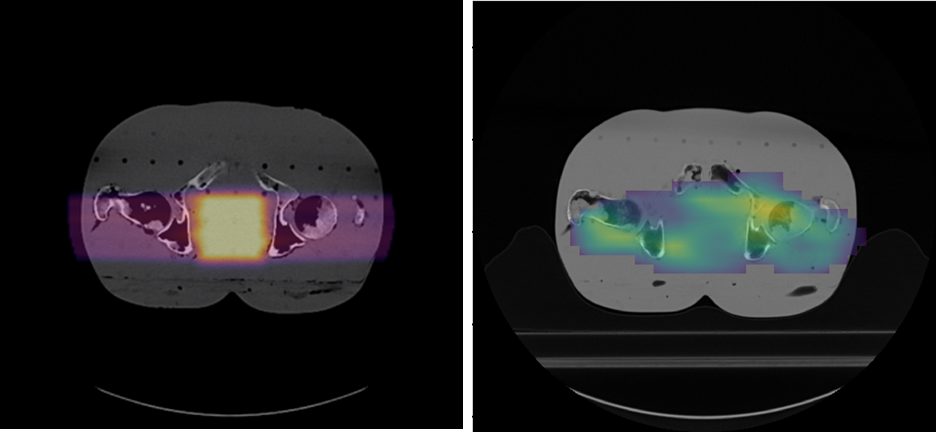}
\caption{
Combined prompt-gamma imaging (PGI) reconstruction overlaid on the CT image of an anthropomorphic phantom. 
The reconstruction integrates prompt-gamma data acquired at gantry angles of $90^{\circ}$ and $270^{\circ}$ during SOBP irradiation of the phantom. The combined reconstruction demonstrates the ability to 
integrate measurements from multiple beam orientations to obtain an overall prompt-gamma emission distribution 
across the full irradiation geometry.
}
\label{fig:combined_pgi}
\end{figure}
In addition to the one-dimensional quantitative analysis, Fig.~\ref{fig:overlay_profile} provides a CT-registered visualization of the reconstructed prompt-gamma emission in axial view. The emission distribution exhibits strong spatial correspondence with the proton beam path and underlying anatomical structures, demonstrating accurate localization of prompt-gamma events within the patient geometry. The reconstructed prompt-gamma hotspot appears along the expected beam trajectory and is consistent with the beam path defined in the treatment plan. This spatial agreement further confirms that the nozzle-mounted quad-camera Compton imaging system can localize prompt-gamma emission relative to anatomical structures. Such CT-registered visualization enables direct anatomical interpretation of the reconstructed prompt-gamma distribution and facilitates comparison with treatment planning information, providing a practical approach for in-room verification of proton beam range and delivery.

This result highlights the capability of the proposed quad-camera Compton imaging system to produce anatomically meaningful prompt-gamma images, which is essential for clinical translation and in vivo proton range verification.

\begin{figure*}[t]
\centering
\includegraphics[width=0.9\textwidth]{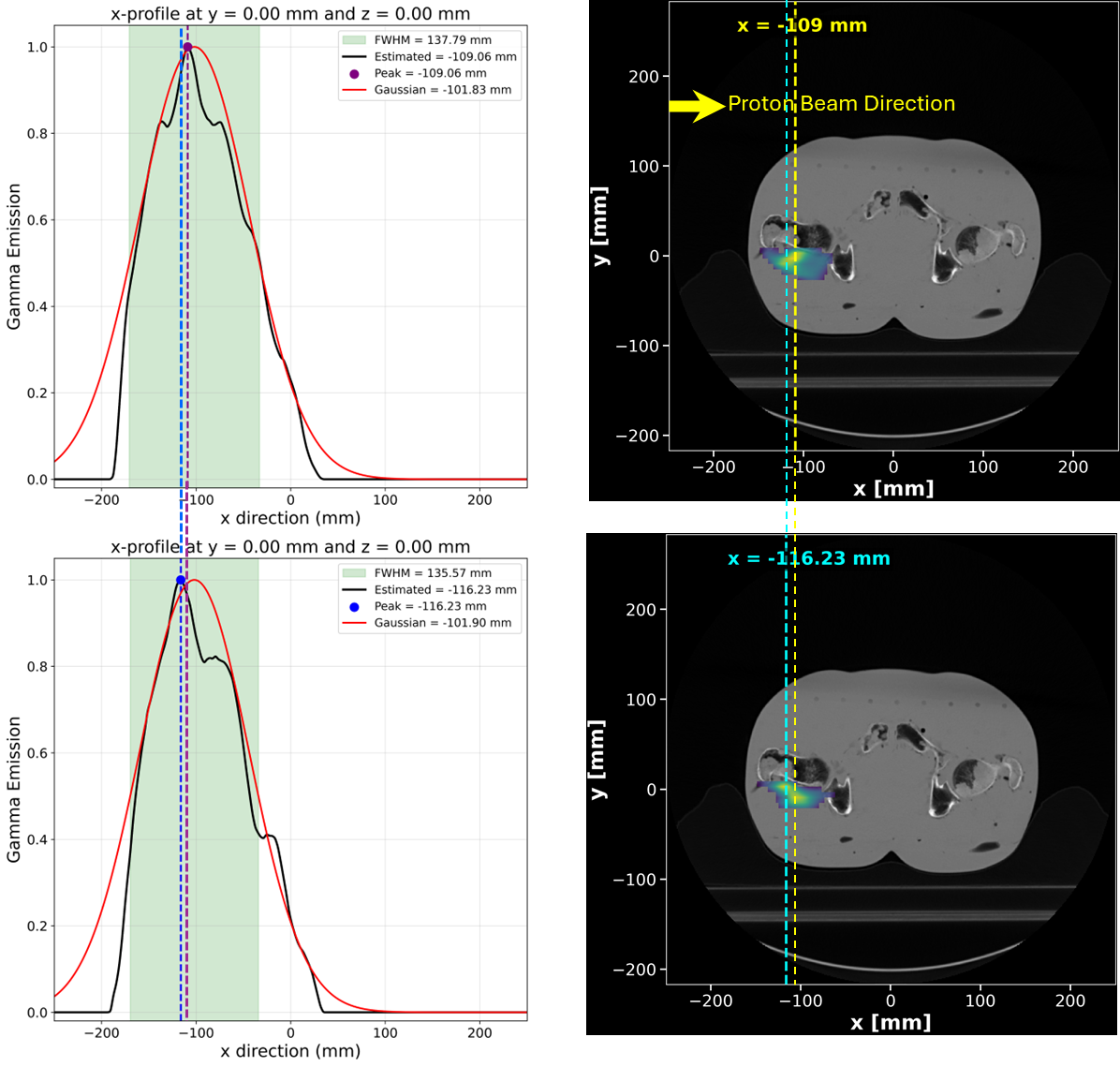}
\caption{
Comparison of reconstructed prompt-gamma emission distributions for controlled proton range shifts of 5 mm (top row) and 10 mm (bottom row) using a 150 MeV single-spot beam and the nozzle-mounted quad-camera Compton imaging system. Left panels: extracted x-profiles of the reconstructed emission at $y = 0$ mm and $z = 0$ mm. The measured prompt-gamma distribution (black) is fitted with a Gaussian model (red), with the shaded region indicating the full width at half maximum (FWHM). The dashed vertical lines denote the estimated centroid positions of the reconstructed emission peaks. Right panels: CT-registered axial views of the reconstructed prompt-gamma emission. The dashed lines indicate the reference proton range and the centroid location of the reconstructed emission peak along the beam axis. Increasing the induced range shift from 5 mm to 10 mm results in a measurable displacement of the reconstructed emission peak along the beam direction, while the transverse emission distribution remains largely unchanged, indicating stable lateral localization.
}
\label{fig:range_shift_detection}
\end{figure*}

\subsection{Angular Dependence: SOBP Beam (G90 vs G270)}

Angular robustness of the quad-camera PGI system was evaluated using a
7.5~Gy SOBP proton beam delivered at gantry angles of $90^\circ$ and
$270^\circ$. Fig.~\ref{fig:gantry_profile_comparison} compares the
lateral prompt-gamma emission profiles reconstructed for the two
gantry orientations. Despite the rotation of the treatment gantry, the
reconstructed emission peaks remain centered near the expected beam
position with similar spatial widths, indicating consistent lateral
localization performance. The measured profiles show comparable
Gaussian fits and FWHM values,
demonstrating stable reconstruction geometry across different treatment
orientations.

In these experiments, the proton beam was delivered using a spread-out Bragg peak (SOBP) configuration with a total prescribed dose of 7.5 Gy. The reconstructed 3D prompt-gamma emission distributions for both gantry angles are shown in Fig.~\ref{fig:AngleEffect}. In both cases, the emission hotspot is clearly localized along the proton beam path. The reconstructed emission exhibits a longitudinal elongation along the beam direction, which reflects the physical distribution of prompt-gamma production during proton transport through the irradiated medium. Prompt gammas are generated continuously as protons undergo nuclear interactions along their path before reaching the distal region of the SOBP. Because the emission originates from an extended region along the beam trajectory rather than a single point, the reconstructed distribution naturally appears elongated along the beam axis.

Measurements were performed for two gantry orientations, G270 and G90, corresponding to the delivery of the proton beam from opposite lateral directions around the treatment isocenter. Despite the change in beam incidence angle relative to the patient and detector system, the reconstructed emission distributions exhibit consistent transverse localization, indicating that the Compton camera mounted on a nozzle maintains a stable geometric response during gantry rotation. These results demonstrate that the system can reliably localize prompt-gamma emission along the proton beam path during clinical beam delivery, supporting its potential for in-room verification of proton beam range across different gantry angles.

To further verify spatial alignment, the reconstructed prompt-gamma distributions were registered with the CT images from the treatment planning dataset, as shown in Fig.~\ref{fig:combined_pgi}. The SOBP irradiation and corresponding treatment plan were generated for the anthropomorphic phantom used in the experiment. The reconstructed emission hotspot follows the expected proton beam trajectory within the phantom, confirming that the prompt-gamma signal accurately tracks the beam path. In this experiment, prompt-gamma reconstructions obtained at gantry angles of $90^{\circ}$ and $270^{\circ}$ were combined to form a single integrated emission map. The resulting distribution demonstrates that prompt-gamma measurements acquired from different gantry orientations can be aggregated to provide a more complete representation of the emission profile. 

This result highlights the capability of the nozzle-mounted quad-camera system to integrate measurements across multiple beam angles while maintaining consistent spatial localization relative to the anatomical CT reference, demonstrating improved robustness during gantry rotation compared with fixed-room prompt-gamma imaging configurations.

\subsection{Sensitivity to Distal Range Shifts}

The sensitivity of the quad-camera PGI system to
proton range variations was evaluated using controlled distal range
perturbations during the delivery of a 150~MeV single-spot proton beam
(80~nA, 60~kMU). Fig.~\ref{fig:range_shift_detection} compares reconstructed
prompt-gamma emission distributions for introduced distal range shifts
of 5~mm and 10~mm. The vertical dashed lines in the figure indicate the location of the maximum value in the two reconstructed PG emissions profiles, highlighting the range shift of the emission hotspot along the beam axis.

A shift of the reconstructed prompt-gamma emission hotspot is observed along the proton beam axis as the applied range perturbation increases. Although the proton beam propagates from left to right, the introduction of additional range-shifting material causes the reconstructed emission hotspot to move upstream (right to left), indicating a reduction in proton penetration depth. As illustrated in the right panels of Fig.~\ref{fig:range_shift_detection}, the peak position shifts from $-109$~mm (yellow line) to $-116$~mm (blue line) as the applied range perturbation increases from $5$~mm (top) to $10$~mm (bottom). This upstream displacement is consistent with the expected shortening of the proton range caused by the added material in the beam path. 

A shift of the reconstructed prompt-gamma emission hotspot is observed along the proton beam axis as the range shift increases. Although the beam travels from left to right, adding range perturbation moves the hotspot upstream (right to left), indicating reduced proton penetration. In Fig.~\ref{fig:range_shift_detection}, the peak position shifts from $-109.06$~mm (yellow line) to $-116.23$~mm (cyan line) when the range perturbation increases from $5$~mm (top) to $10$~mm (bottom), consistent with the added material shortening the proton range. While the measured shift of approximately $7$~mm is larger than the expected $5$~mm shift, this increase remains significant and the difference is consistent with the spatial resolution of the reconstruction algorithm.

To quantitatively evaluate how the applied range shift affects the reconstructed emission distribution, one-dimensional prompt-gamma intensity profiles were extracted along the $x$ direction at $y = 0$~mm and $z = 0$~mm, as shown in the left panels of Fig.~\ref{fig:range_shift_detection}. The measured emission profiles are represented by the black curves, while the red curves correspond to Gaussian fits used to estimate the peak position and spatial spread of the distribution. The shaded green region indicates the FWHM of the fitted profile. The dashed vertical lines mark the estimated peak locations of the reconstructed emission hotspot. Comparing the two measurements shows that the peak position shifts upstream as the range perturbation increases, demonstrating that the applied range shift directly affects the reconstructed prompt-gamma emission distribution. This upstream displacement is consistent with the proton beam stopping earlier in the phantom.

These results demonstrate a strong sensitivity of the nozzle-mounted quad-Compton camera
PGI system to millimeter-scale distal range variations under clinical beam delivery conditions. The system reliably detect and localize range shifts, highlighting its robustness and effectiveness for PGI range verification and demonstrating improved three-dimensional localization of prompt-gamma emission along the proton beam path. Importantly, because prompt gamma emission is strongly correlated with
the location of proton nuclear interactions along the beam path, the
observed displacement of the emission maximum provides a direct and experimentally validated indicator for changes in the distal proton range and therefore for the
location of the Bragg peak. This capability represents a key step toward real-time
\textit{in vivo} verification of proton beam range during clinical treatment delivery.

\FloatBarrier
\section{Conclusion}
\label{sec:Conclusion}
This work presents the first clinical integration of a nozzle-mounted Compton camera prompt-gamma imaging (PGI) system for proton therapy and demonstrates its capability to detect millimeter-scale range variations under routine beam delivery conditions. The system operated stably during clinical irradiations without disrupting the treatment geometry or workflow, while enabling real-time multi-camera acquisition and in-phantom 3D emission reconstruction.

Controlled distal range perturbations produced consistent and measurable displacements in the prompt-gamma emission profile, confirming quantitative sensitivity to proton range changes. In addition, stable performance across multiple beam energies, gantry angles, and delivery modalities demonstrates the robustness of the PGI system under varying clinical conditions. Furthermore, the localization of prompt-gamma emission along the proton beam path validates the system’s capability for three-dimensional detection and angular consistency. Collectively, these results establish the feasibility of \textit{in vivo} proton range verification using a clinically deployable quad-camera prompt-gamma imaging system.

By enabling direct monitoring of proton range during treatment delivery, this approach provides a practical pathway toward reducing range uncertainties and improving the safety and precision of proton therapy. Future work will focus on integrating machine learning to further enhance reconstruction accuracy and spatial localization, in addition to dose-rate scaling, patient-specific validation, and workflow optimization to support full clinical translation.

\section*{Acknowledgment}
This work was supported by the National Institutes of Health (NIH), 
National Cancer Institute (NCI), under Grant R01CA279013,
“3-Dimensional Prompt Gamma Imaging for Online Proton Beam Dose Verification.”

We acknowledge the UMBC High Performance Computing Facility (HPCF, \texttt{hpcf.umbc.edu}) and the financial contributions from the NIH, NSF, CIRC, and UMBC that supported this work.

\bibliographystyle{IEEEtran}
\bibliography{references}
\end{document}